\documentclass[twocolumn,pra,showpacs,amsmath,amssymb]{revtex4-1}
\usepackage{graphicx}
\usepackage{bm}
\usepackage{epstopdf}
\usepackage{epsfig}
\usepackage{dsfont}
\usepackage{amssymb}
\begin{document}

\title{Enlarging the notion of additivity of resource quantifiers}
\author{L. F. Melo, Thiago Melo, and Fernando Parisio}
\email[]{parisio@df.ufpe.br}
\affiliation{Departamento de
F\'{\i}sica, Universidade Federal de Pernambuco, Recife, Pernambuco
50670-901 Brazil}

\begin{abstract}
Whenever a physical quantity becomes essential to the realization of useful tasks, it is desirable to define proper measures or monotones to quantify it. 
In quantum mechanics, coherence, entanglement, and Bell nonlocality are examples of such quantities. Given a quantum state $\varrho$ and a quantifier 
${\cal E}(\varrho)$, both arbitrary, it is a hard task to determine ${\cal E}(\varrho^{\otimes N})$.
However, if the figure of merit $\cal{E}$ turns out to be additive, we simply have ${\cal E}(\varrho^{\otimes N})=N e$, with $e={\cal E}(\varrho)$.
In this work we generalize this useful notion through the inner product ${\cal E}(\varrho^{\otimes N}) =
\vec{N}\cdot \vec{e}$, where $\vec{e}=({\cal E}(\varrho^{\otimes i_1}), {\cal E}(\varrho^{\otimes i_2}),\dots,{\cal E}(\varrho^{\otimes i_q}) )$ is a vector whose $q$ entries
are the figure of merit under study calculated for some numbers of copies {\it smaller} than $N$ ($1 \le i_1<i_2<\dots <i_q<N$), where $\vec{N}=(N_{i_1}, N_{i_2}, \dots ,N_{i_q})$,
is a string of numbers that depends only on $N$ and on the set of integers $\{ {i_j}\}$. We show that the one shot distillable
entanglement of certain spherically symmetric states can be quantitatively approximated by such an augmented additivity. 
\end{abstract}
\maketitle

\section{Introduction}
\label{intro}
Many quantum mechanical tasks of either foundational or practical relevance, involving a particular state $\varrho$,
require the preparation of a large number of copies of this state to be accomplished. In some situations,
it turns out that employing $\varrho^{\otimes N}$ and global operations over the large Hilbert-Schmidt space it inhabits
proves to be advantageous in comparison with the isolated manipulation of $N$ copies of $\varrho$. This is an instance
of the nonadditivity of the figure of merit that embodies the ability to execute the task under consideration.

Additivity is a special property satisfied only by a handful of measures or monotones of quantum resources \cite{comm0}, 
as for instance, the logarithmic negativity \cite{vidal} and squashed entanglement \cite{squashed}, for non separability \cite{horodecki}. For these
special quantifiers ${\cal E}(\varrho^{\otimes N})=N{\cal E}(\varrho)$.
Thus, additivity renders an otherwise hard problem trivial: the evaluation of the amount of resources contained
in $N$ copies of a given state $\varrho$, ${\cal E}(\varrho^{\otimes N})$, once we know ${\cal E}(\varrho)$.
We remark  that a stronger notion of additivity is ${\cal E}(\varrho\otimes\sigma)={\cal E}(\varrho)+{\cal E}(\sigma)$, for
all pairs of states $\varrho$ and $\sigma$. In this manuscript, whenever we refer to additivity, we mean the weaker
condition.

The majority of quantum-resource quantifiers, however, is nonadditive. Examples regarding nonseparability are the entanglement of formation \cite{formation, hastings} and the geometric measure of entanglement \cite{geom}, among others \cite{nonadd, dafa, dafa2}.  An extreme example of such a behavior is the phenomenon of superactivation, for which, given a particular quantifier ${\cal E}$, we may find 
states $\varrho$, such that ${\cal E}(\varrho)=0$ and ${\cal E}(\varrho^{\otimes N})>0$, for some $N>1$, 
as is the case of the distillable entanglement \cite{superact,watrous}. 

The technical difficulty to determine ${\cal E}(\varrho^{\otimes N})$ is an important motivation behind the search for asymptotic limits, 
which are useful whenever one can assume that the number of states is large enough to make the limit
$N\rightarrow \infty$ an acceptable approximation.
However, often, the asymptotic regime dominates only for a unpractically high number of copies \cite{oneshot,natcomm,ieee}. 
Therefore, one cannot always evade the question of evaluating ${\cal E}$ for a large, but finite number of copies. 

An alternative attempt to deal with the problem comes from the concept of scalable quantifiers, introduced in \cite{scal1}, which is summarized in the next section. The remanining sections of this work explore a generalization of the additivity notion, stemming from scalability, which can prove useful in the evaluation of some relevant figures of merit in the large $N$ limit.
\section{Preliminaries: Scalable quantum functions}
Let us initially formalize the notion of scalability \cite{scal1}. 
Consider an arbitrary function ${\cal E}$ with argument $\varrho^{\otimes N}$, ${\cal E}: {\cal B(H)}^{\otimes N} \mapsto \mathds{R}_{+}$. In practice, the quantity ${\cal E}(\varrho^{\otimes N})$ can be a measure or monotone of a quantum resource, or some cost function, for instance. Now, suppose we either know by definition or find out that this function 
is completely (or approximately) determined by $q$ real numbers, namely, $e_{i_1}={\cal E}(\varrho^{\otimes i_1})$, $e_{i_2}={\cal E}(\varrho^{\otimes i_2}), \dots, e_{i_q}={\cal E}(\varrho^{\otimes i_q})$, $ i_j<N$, $j=1, \dots, q$.
That is, once we know the value of ${\cal E}$ for some smaller number of copies, ${\cal E}(\varrho^{\otimes N})$ is determined:
\begin{equation}
\label{def0}
{\cal E}(\varrho^{\otimes N})=E^{(N)}(\vec{e}),
\end{equation}
where, for short, we collect the non-negative numbers $e_{i_j}$ as the components of a vector in the positive hyperoctant of $\mathds{R}^q$, 
$$ \vec{e}\equiv (e_{i_1}, e_{i_2}, \dots, e_{i_q}).$$

One of the main results of \cite{scal1} is that the possible functional forms of $E^{(N)}(\vec{e})$ must satisfy the constraint
\begin{equation}
\label{recur2}
E^{(N)}(\vec{e})=E^{(N/K)}\left(E^{(i_1K)}(\vec{e}), E^{(i_2K)}(\vec{e}), \dots, E^{(i_qK)}(\vec{e})\right),
\end{equation}
for $N=a^n$ and $K=a^k$, with $n,k \in \{0,1,2,\dots\}=\mathds{N}$, $k\le n$. So $N$ and $K$ take values on 
$$\mathds{P}_{a}=\{1,a, a^2, \dots\},$$  
the set of all non-negative integer powers of an arbitrary positive integer $a$.

The restriction embodied by (\ref{recur2}) stems from the fact that, while the left-hand side of (\ref{def0}) automatically comply with the structure
of tensor products (the domain of ${\cal E}$ is a Hilbert-Schmidt space), for the right-hand side, this compliance must be imposed on the functional form of $E$,
which is defined on $\mathds{R}^q$.
More specifically, ${\cal E}(\varrho^{\otimes N})={\cal E}(\sigma^{\otimes (N/K)})$ is a tautology, whenever
\begin{equation}
\label{eq}
\sigma^{\otimes (N/K)}=\varrho^{\otimes N}, \;\mbox{with}\; \sigma\equiv\varrho^{\otimes K}.
\end{equation}
Equation (\ref{recur2}) guarantees that $E$ will not violate this rule.
A figure of merit ${\cal E}(\varrho^{\otimes N})$ that can be expressed through (\ref{def0}) and satisfying the above
consistency condition is referred to as a $q$-scalable ($q$-S) function. 

In the simplest case, $q=1$, the constraint relation reads
\begin{equation}
\nonumber
E^{(N)}(e_1)=E^{(N/K)}\left(E^{(K)}(e_1)\right).
\end{equation}   
Note that, while ${\cal E}: {\cal B(H)}^{\otimes N} \mapsto \mathds{R}_{+}$, we have
$E:\mathds{R}_{+}^{q} \mapsto  \mathds{R}_{+}$, so that, typically, 
the domain of the latter has a dimension which is much lower than that of the former.
In general, the functional dependencies that satisfy the constraint imposed by relation (\ref{recur2}), may be nonlinear, the $\ell_1$-norm of coherence \cite{coh} being a simple but nontrivial example of a 1-scalable function \cite{scal2}. 
\section{Augmented  additivity}
Inspired by the additivity property as expressed by $E^{(N)}(e)=N e$ (which is the simplest 1-S function), we will investigate $q$-scalable functions which, in addition, present a linear dependence on the components of $\vec{e}$:
\begin{equation}
{\cal E}(\varrho^{\otimes N}) =E^{(N)}(\vec{e})=\vec{N}\cdot \vec{e}.
\label{qadd}
\end{equation}
This augmented notion of additivity, here referred to as $q$-additivity, along with the general constraint (\ref{recur2}), give rise to explicit recurrence relations that must be satisfied by the entries of $\vec{N}$, as will be shown in the next subsection, for $\vec{e}=(e_{i_1},e_{i_2}) \in \mathds{R}^2$ ($q=2$). For $q=1$ the only possible linear dependence is that of usual additivity, $E^{(N)}(e)=N e$.
\subsection{2-S Recurrence Relations}\label{section:2s}
Here we will be concerned with functions ${\cal E}(\varrho^{\otimes N})=E^{(N)}(e,f)$, where, to simplify the notation, we set $e_{i_1}=e$ and $e_{i_2}=f$. 
Although this case has been considered in \cite{scal1} as a first order approximation for general 2-scalable functions, our procedure here is distinct, and will serve 
as a basis for the general results to be developed. Also, the strict linearity hypothesis leads to closed analytical asymptotic results, presented in the end of this section.

Assuming 2-additivity, Eq. (\ref{qadd}), we write
\begin{equation}
E^{(N)}(e,f)={\cal A}_Ne+{\cal B}_Nf,
\end{equation}
with $\vec{N}=({\cal A}_N,{\cal B}_N)$, while the scalability condition, Eq.(\ref{recur2}), simplifies to
\begin{eqnarray}
E^{(N)}(\vec{e})=E^{(N/K)}\left[E^{(K)}(e,f),E^{(aK)}(e,f)\right].
\end{eqnarray}
Gathering these two equations together gives
\begin{eqnarray}
{\cal A}_Ne + {\cal B}_Nf &=& {\cal A}_{N/K}E^{(K)}(e,f) +  {\cal B}_{N/K}E^{(aK)}(e,f) \label{2spresentation} \nonumber\\
&=&  {\cal A}_{N/K} \left[{\cal A}_Ke + {\cal B}_Kf\right] \nonumber  \\
&+& {\cal B}_{N/K}\left[{\cal A}_{aK}e + {\cal B}_{aK}f\right].
\end{eqnarray}
To determine the coefficients it is sufficient to set $K=a$, such that ${\cal A}_a=0$ and ${\cal B}_a=1$. Combining the terms proportional to $e$ and $f$ and using the fact that these variables are assumed to be independent, we get the following coupled recurrence relations:
\begin{eqnarray} 
A_n &=& x\text{ }B_{n-1}, \nonumber \\ 
B_n &=& A_{n-1} + y\text{ }B_{n-1}, \nonumber 
\end{eqnarray}
where we used ${\cal A}_{a^2}=x$, ${\cal B}_{a^2}=y$, and defined ${\cal A}_N = A_n$ and ${\cal B}_N = B_n$, with $N=a^n$.
By decoupling the above equations we get linear, homogeneous, second order recurrence relations with constant coefficients:
\begin{eqnarray}
B_n = y\text{ }B_{n-1}+x\text{ }B_{n-2}\label{recurrence1}
\end{eqnarray}
Similarly for $A_n$. The solution of the previous recurrence relation is given by the generalized hybrid Fibonacci polynomials $B_n = F_n(x,y)$ \cite{fibonacci} of two variables, $x$ and $y$, and degree $n$:
\begin{equation}
F_n(x,y) = \sum_{k=0}^{\lfloor{(n-1)/2}\rfloor} \left(\begin{array}{c} n-1-k \\ k \end{array}\right)x^{n-1-2k} y^k.
\label{fconstruction}
\end{equation}
These polynomials are a general form of the famous Fibonacci Numbers, which are obtained by setting $x=y=1$.

Therefore, if $E$ is a 2-additive function such that $E(\varrho^{\otimes a^n})$ depends only on $e = {\cal E}(\varrho)$, $f = {\cal E}(\varrho^{\otimes a})$ and $n$, where $E(\varrho^{\otimes a^2}) = xe+yf $, with $x$ and $y$ known, then, we must have:
\begin{eqnarray}
\label{2sfibonacci}
E^{(a^n)}(\vec{e}) =\left[xF_{n-1}(x,y) , F_n(x,y)\right]\cdot(e,f)\nonumber\\
= xF_{n-1}(x,y) e + F_n(x,y)f,
\end{eqnarray}
for {\it arbitrary} $n$.
Note that, in order to use this expression, in addition to the numeric values of ${\cal E}(\varrho)$ and ${\cal E}(\varrho^{\otimes a})$ ($e$ and $f$ respectively), it is necessary to know how the next-order quantity ${\cal E}(\varrho^{\otimes a^2})=E^{(a^2)}(e,f)$ depends on $e$ and $f$, that is, the coefficients $x$ and $y$ in $E^{(a^2)}(e,f)=xe+yf$. 
The value of ${\cal E}(\varrho^{\otimes a^3})$ and ${\cal E}(\varrho^{\otimes a^4})$, for instance, are given by
\begin{eqnarray}
\nonumber
E^{(a^3)}(e,f)&=&xye+\left(x+y^{2}\right)f, \\
\nonumber
E^{(a^4)}(e,f)&=&x\left(x+y^{2}\right)e+y\left(2x+y^{2}\right)f.
\end{eqnarray}

The general, solution is given by the explicit form of the Fibonacci Polynomials \cite{fibonacci} and reads
\begin{eqnarray}
A_n &=& \frac{x}{\sqrt{\cal Z}} \left[\left(\frac{y+\sqrt{\cal Z}}{2}\right)^{n-1} - \left(\frac{y-\sqrt{\cal Z}}{2}\right)^{n-1} \right], \label{An} \\ 
B_n &=& \frac{1}{\sqrt{\cal Z}} \left[\left(\frac{y+\sqrt{\cal Z}}{2}\right)^{n} - \left(\frac{y-\sqrt{\cal Z}}{2}\right)^{n} \right], \label{Bn}
\end{eqnarray}
where ${\cal Z}=4x+y^2$. Note that not all values of $x$ and $y$ are permitted, the inequality $y^2\ge-4x$ must be satisfied. Several such inequalities apear in the formalism of scalable
quantifiers and constitute necessary conditions for the application of the theory.

We now proceed to a change of variables that will be useful to formalize the generalization of these results to $q$-additivity. We replace the variables $x$ and $y$ with $\nu_1$ and $\nu_2$, through:
\begin{equation}
\frac{y+\sqrt{\cal Z}}{2} = a^{\nu_1},\;\;	\frac{y-\sqrt{\cal Z}}{2} = a^{\nu_2}, \label{change2} 
\end{equation}
where $\sqrt{\cal Z}=\sqrt{4x+y^2}=a^{\nu_1}-a^{\nu_2}$. Since $a^n=N$, by substituting (\ref{change2}) into (\ref{An}) and (\ref{Bn}) we get final expressions in terms of the number of copies $N$:
\begin{eqnarray}
{\cal A}_N &=& \frac{-a^{\nu_1+\nu_2}}{a^{\nu_1}-a^{\nu_2}} \left[\left(\frac{N}{a}\right)^{\nu_1}-\left(\frac{N}{a}\right)^{\nu_2}\right] \label{A(N)}, \\
{\cal B}_N &=& \left(\frac{N^{\nu_1}-N^{\nu_2}}{a^{\nu_1}-a^{\nu_2}}\right) \label{B(N)}.
\end{eqnarray}
We, thus, can state that any 2-additive function must be of the form:
\begin{equation}
E^{(N)}(e,f) = \left(\frac{N^{\nu_2}a^{\nu_1}- N^{\nu_1}a^{\nu_2}}{a^{\nu_1}-a^{\nu_2}}\right)e + \left(\frac{N^{\nu_1}-N^{\nu_2}}{a^{\nu_1}-a^{\nu_2}}\right) f,  \label{general2s}
\end{equation}
where
\begin{equation}
\nu_1 = \log_a{\left(\frac{y+\sqrt{\cal Z}}{2}\right)} ,\;\;\nu_2 = \log_a{\left(\frac{y-\sqrt{\cal Z}}{2}\right)}. \label{n1,n2}
\end{equation}
The change of variables (\ref{change2}) determines $x$ and $y$ for a 2-additive function (\ref{general2s}). By inverting the expressions (\ref{change2}) we get:
\begin{equation}
\left\{\begin{array}{rll}
x &=& - a^{\nu_1 + \nu_2} \\
y &=& a^{\nu_1} + a^{\nu_2} \label{x,y}
\end{array}\right.
\end{equation}
Thus, $x$ is \textit{always} a negative number. Then, because of equations (\ref{An}) and (\ref{Bn}), these coefficients must satisfy $y^2\geq 4|x|$ for (\ref{general2s}) to be a physically acceptable quantity. A different constraint can be obtained as follows. Since ${\cal E}(\varrho^{\otimes N}) \ge {\cal E}(\varrho^{\otimes M})$, for $N \ge M$,
we must have $x e+y f\ge f$, i. e., $f/e\ge |x|/(y-1)$. If these inequalities are not observed, the figure of merit cannot possibly be 2-additive.

Note that expression (\ref{general2s}) is symmetric under exchange of the exponents $\nu_1$ and $\nu_2$.  The regularized function $\lim_{N \rightarrow \infty}(E^{(N)}(\vec{e})/N)$ vanishes for $\nu_1,\nu_2 < 1$, while it diverges whenever either $\nu_1>1$ or $\nu_2 >1$. In both cases we assumed $\nu_1\ne1$ and $\nu_2\ne 1$. However, if either $\nu_1=1$ (with $\nu_2 <1$) or $\nu_2=1$ (with $\nu_1<1$), the regularized function turns out to be finite and non-zero. Given the symmetry $\nu_1 \leftrightarrow \nu_2$, without loss of generality, we take $\nu_1=1$ and $\nu_2\equiv\nu<1$. In this case. it is easy to verify that
\begin{equation}
\lim_{N \rightarrow \infty} \frac{E^{(N)}(e,f)}{N} = \frac{f-a^{\nu}e}{a-a^{\nu}}, \nonumber
\end{equation}
where $\nu<1$ and $f>a^{\nu}e$. So, if we know that a quantity of interest is 2-additive and, also, has a finite asymptotic value per copy, then, the above relation must hold. In fact, if this asymptotic value is known in advance, then the value of $f$ can be determined from it and from $e$.

The only remaining case is $\nu_1=\nu_2$, which leads to an indefinite limit of the kind ``$0/0$''. By setting $\nu_1=\nu $ and $\nu_2=\nu+ \delta$ and expanding to first order for small $\delta$ we get:
\begin{equation}
E^{(N)}(e,f) =-N^{\nu}\left(\log_a N-1 \right)e+\frac{N^{\nu}}{a^{\nu}} \log_a N f.
\end{equation}
In particular, for $\nu_1=\nu_2=1$, $E^{(N)}(e,f)/N$ diverges logarithmically as $N\rightarrow \infty$.

\section{Generalization: $q$-additvity}
\label{qadd}

We now set to obtain the general constraints that must be satisfied by $q$-additive functions, for arbitrary $q$, that we express as
\begin{equation}
\label{qadd2}
E^{(N)}(\vec{e})=\vec{N}\cdot\vec{e}=\sum_j  \eta^{i_j}(N)e_{i_j}.
\end{equation}
In this case, the general scalability constraint (\ref{recur2}) reads
\begin{eqnarray}
\nonumber
E^{(N)}(\vec{e})=\sum_{\ell=1}^{q} \eta^{i_\ell}(N/K)E^{(i_\ell K)}(\vec{e})\\
\nonumber
= \sum_{\ell=1}^{q} \eta^{i_\ell}(N/K)\; \sum_{j=1}^{q} \eta^{i_j}(i_\ell K)e_{i_j}\\
= \sum_{j=1}^{q} \left(  \sum_{\ell=1}^{q}\eta^{i_j}(i_\ell K) \eta^{i_\ell}(N/K) \right)e_{i_j}.
\label{condlin}
\end{eqnarray}
Comparing equations (\ref{qadd2}) and (\ref{condlin}) we get
\begin{equation}
\eta^{i_j}(N)= \sum_{\ell=1}^{q} \eta^{i_j}(i_\ell K) \eta^{i_\ell}(N/K). \label{recurlin}
\end{equation}
Therefore, expression (\ref{qadd2}) represents a $q$-additive function whenever (\ref{recurlin}) is satisfied. In the simplest case we have $q=1$ with $i_1=1$ and $\eta_{1}(N)=\eta_{1}(K) \eta_{1}(N/K)$. Again, we remark that to obtain the general solution it suffices to set $K=a$.

Let us simplify the notation using $\eta^{i_\ell}(N) = \eta^\ell_n$ ($N=a^n$), then:
\begin{equation}
\eta^{j}_n= \sum_{\ell=1}^{q} \eta^{j}_\ell \eta^\ell_{n-1}. \label{beforeexpressionforA}
\end{equation}
Note that, by definition, $\eta^1_\ell=0$ except for $\ell=q$ (denoted by $x$ in previous section) and $\eta^2_\ell=0$ except for $\ell=1$ ($a$ copies) and for $\ell=q$ 
(denoted by $y$ in previous section). Following the same reasoning we may generally write:
\begin{equation}
\eta^j_\ell = \underbrace{\delta^{j-1}_\ell}_\text{for $\ell<q$} + \delta^q_\ell \eta^j_q. \label{expressionforA}
\end{equation}
This relation states that every coefficient is null except if it satisfies the consistency conditions $E^{(1)}=e_1$, $E^{(a)}=e_2$, etc, up to $\ell<q$. In addition, for $a^q$ copies we define new values $\eta^1_q(a^q)=x$, $\eta^2_q(a^q)=y$, etc. That is why the Kronecker delta symbols are labeled for an index $\ell<q$ or for $\ell=q$. Substituting (\ref{expressionforA}) into (\ref{beforeexpressionforA}) we get:
\begin{eqnarray}
\eta^{j}_n &=& \sum_{\ell=1}^{q} (\underbrace{\delta^{j-1}_\ell}_\text{for $\ell<q$} + \delta^q_\ell \eta^j_q)\; \eta^\ell_{n-1} \nonumber \\
&=& \eta^{j-1}_{n-1} + \eta^q_{n-1}\eta^j_q \label{general1ºsystem}
\end{eqnarray}
Expression (\ref{expressionforA}) can be put into the following matrix form:
\begin{equation}
\left(\begin{matrix}
\eta^1_n \\
\vdots  \\
\vdots\\
\eta^q_n  \\
\end{matrix} \right) \text{  } = \text{  }\left(\begin{matrix}
0 & 0 & \dots & 0& 0 & \eta^1_q \\
1 & 0 &\dots & 0& 0 & \eta^2_q \\
\vdots & \vdots & \ddots & \vdots &\vdots&\vdots&  \\
0 & 0 & \dots & 1& 0 & \eta^{q-1}_q \\
0 & 0 &\dots & 0& 1 & \eta^q_q \\
\end{matrix} \right) \text{  }\left(\begin{matrix}
\eta^1_{n-1}  \\
\vdots   \\
\vdots\\
\eta^q_{n-1}  \\
\end{matrix} \right) \label{qmatrix}
\end{equation}
Note that the first line is null except for the last entry $\eta^1_q$, while the following lines compose a $(q-1)\times(q-1)$ identity matrix on the left-down block and the last column is given by the numbers $\eta^m_q$. Let us denote this matrix by $ \mathcal{Q}$:
\begin{equation}
\mathcal{\eta}_n = \mathcal{Q}\mathcal{\eta}_{n-1} \label{Qmatrix}.
\end{equation}
It is instructive to test this matrix approach in the 2-additive case:
	\begin{equation}
	\left(\begin{matrix}
	\eta^1_n \\
	\eta^2_n  \\
	\end{matrix} \right) \text{  } = \text{  }\left(\begin{matrix}
	\eta^1_1 & \eta^1_2 \\
	\eta^2_1 & \eta^2_2  \\
	\end{matrix} \right) \text{  }\left(\begin{matrix}
	\eta^1_{n-1}  \\
	\eta^2_{n-1}   \\
	\end{matrix} \right) \text{ }=\text{ } \left(\begin{matrix}
	0 & x \\
	1 & y \\
	\end{matrix} \right) \text{  }\left(\begin{matrix}
	\eta^1_{n-1}  \\
	\eta^2_{n-1}   \\
	\end{matrix} \right),\label{fibonaccimatrix}
	\end{equation}
which indeed leads to the generalized hybrid Fibonacci polynomials (\ref{fconstruction}).

The advantage of this formalism, especially for larger values of $q$, is that one can diagonalize $\mathcal{Q}$ through the operation $\mathcal{\eta}_n = D^{-1}\mathcal{\zeta}_n$ ($D$ is $q \times q$): $D^{-1} \zeta_n = \mathcal{Q} D^{-1} \zeta_{n-1}$, $D D^{-1}\zeta_n = D\mathcal{Q} D^{-1} \zeta_{n-1}$,
where $[D\mathcal{Q}D^{-1}]_{kl} = \lambda_k \delta_{kl}$ is the diagonalized matrix with $\lambda_k$ being the eigenvalues. In this basis the solution of the $q$-additivity problem is multiplicative:
\begin{equation}
\zeta^k_n = (\lambda_k)^n,
\end{equation}
and as $\eta_n$ is the result of a matrix operation $D^{-1}$ on the vector $\zeta_n$, we can expand it in a sum of coefficients $C^m_k$. These coefficients will be the solutions of  determined systems of equations related to expression (\ref{expressionforA}).
\begin{equation}
\eta^m_n = \sum_{k=1}^q C^m_k \zeta^k_n = \sum_{k=1}^q C^m_k (\lambda_k)^n 
\end{equation}
\subsection{Eigenvalues and Coefficients Equations}
The eigenvalues of the $\mathcal{Q}$-matrix in (\ref{qmatrix}) are the key ingredients of the method. Note that the eigenvalues of (\ref{fibonaccimatrix}) are exactly the values we used in the change of variables (\ref{change2}). So we can generalize this step to:
\begin{equation}
\nu_k = \log_a{\lambda_k}, \label{generalchange}
\end{equation}
where $\lambda_i$ is the $i$th eigenvalue of $\mathcal{Q}$ (So $i$ goes from 1 to $q$), therefore being a root of the $q$-order characteristic polynomial. Then the general solution, by making the change of variables (\ref{generalchange}) is:
\begin{eqnarray}
\nonumber
\eta^m_n = \sum_{k=1}^q C^m_k(\lambda_1,\dots,\lambda_q) (\lambda_k)^n \text{ }\text{ }\text{ }\text{ }\text{ } \\
\Rightarrow \text{ }\text{ }\text{ }\text{ } \eta^m(N) = \sum_{k=1}^q C^m_k(a^{\nu_1},\dots,a^{\nu_q}) N^{\nu_k}, \label{coefficientsC}
\end{eqnarray}
where the coefficients $C^m_k$ are functions of the eigenvalues of $\mathcal{Q}$ and must satisfy the defined boundary conditions, $\eta^m_l=\delta^{m-1}_l$, for $l=0,\dots,q-1$. As each coefficient $\eta^m(a^l)$ is a sum (\ref{coefficientsC}) with coefficients $C^m_k(a^{\nu_1},\dots,a^{\nu_q})$ these boundary conditions give rise to $q$ systems of $q$ coupled, but linear equations to be solved. After one solves these systems of equations the problem is finished:
\begin{eqnarray}
E^{(N)}(\vec{e}) &=& \sum_{m=1}^q \sum_{k=1}^q C^m_k N^{\nu_k} e_m  \nonumber \\
\sum_{k=1}^q C^m_k a^{l\nu_k} &=& \delta^{m-1}_l \text{ }\text{ }\text{ }, \text{ }\text{ }\text{ } l=0,...,q-1. \nonumber 
\end{eqnarray}

Let us resume the 2-additive case, using the matrix formalism. The eigenvalues of:
\begin{equation}
\mathcal{Q} = \left(\begin{matrix}
0 & x \\
1 & y \\
\end{matrix} \right)
\end{equation}
are $\lambda_1 =(y+\sqrt{\cal Z})/2$ and $\lambda_2 =(y-\sqrt{\cal Z})/2$. Then:
\begin{equation}
\eta^m(N) = C^m_1 N^{\nu_1} + C^m_2 N^{\nu_2} ,
\end{equation}
where $\nu_1=\log_a \lambda_1$ and $\nu_2=\log_a \lambda_2$, with boundary conditions $\eta^1(1)=1, \eta^1(a)=0$ and $\eta^2(1)=0, \eta^2(a)=1$. We get two systems of equations to solve:
\begin{eqnarray}
\left\{\begin{array}{rll}
C^1_1 + C^1_2 &=& 1 \\
C^1_1 a^{\nu_1} + C^1_2 a^{\nu_2} &=& 0 \nonumber
\end{array}\right.
\end{eqnarray}
implying
$$C^1_1(a^{\nu_1},a^{\nu_2}) = \frac{-a^{\nu_2}}{a^{\nu_1}-a^{\nu_2}} = 1- C^1_2(a^{\nu_1},a^{\nu_2}),$$
and
\begin{eqnarray}
\left\{\begin{array}{rll}
C^2_1 + C^2_2 &=& 0 \\
C^2_1 a^{\nu_1} + C^2_2 a^{\nu_2} &=& 1 \nonumber
\end{array}\right. 
\end{eqnarray}
with solution 
$$C^2_1(a^{\nu_1},a^{\nu_2})= \frac{1}{a^{\nu_1}-a^{\nu_2}}=-C^2_2(a^{\nu_1},a^{\nu_2}).$$
These lead, exactly, to the results we have found in (\ref{A(N)}) and (\ref{B(N)}).
\section{3-additivity}
The next simpler case is that of 3-additivity, for which we have:
\begin{equation}
\mathcal{Q} = \left(\begin{matrix}
0 & 0 & x \\
1 & 0 & y \\
0 & 1 & z 
\end{matrix} \right).
\end{equation}
One can find the three eigenvalues of this matrix, change variables $\nu_k=\log_a \lambda_k$, $k=1,2,3$, and write:
\begin{equation}
\eta^m(N) = C^m_1N^{\nu_1} + C^m_2N^{\nu_2} + C^m_3N^{\nu_3} 
\end{equation}
Now we have boundary conditions $\eta^1(1)=1, \eta^1(a)=\eta^1(a^2)=0$, $\eta^2(1)=\eta^2(a^2)=0, \eta^2(a)=1$ and $\eta^3(1)=\eta^3(a^2)=0, \eta^3(a^2)=1$. So, there are three systems of equations to be solved:
\begin{eqnarray}
\left\{\begin{array}{rll}
C^m_1 + C^m_2 + C^m_3 &=& \delta^{m-1}_0 \nonumber \\
C^m_1 a^{\nu_1} + C^m_2 a^{\nu_2} + C^m_3 a^{\nu_3} &=& \delta^{m-1}_1 \nonumber \\
C^m_1 a^{2\nu_1} + C^m_2 a^{2\nu_2} + C^m_3 a^{2\nu_3} &=& \delta^{m-1}_2 \nonumber
\end{array}\right.
\end{eqnarray}
These systems can be easily dealt with through some software capable of symbolic manipulations, like mathematica. The nine coefficients we obtained read:
\begin{eqnarray}
	C^1_1 &=& \frac{a^{\nu_2+\nu_3}}{(a^{\nu_1}-a^{\nu_2})(a^{\nu_1}-a^{\nu_3})}, \nonumber\\
	C^1_2 &=& \frac{a^{\nu_1+\nu_3}}{(a^{\nu_2}-a^{\nu_1})(a^{\nu_2}-a^{\nu_3})}, \nonumber\\
	C^1_3 &=& \frac{a^{\nu_1+\nu_2}}{(a^{\nu_3}-a^{\nu_1})(a^{\nu_3}-a^{\nu_2})}\nonumber\\
	C^2_1 &=& \frac{-(a^{\nu_2}+a^{\nu_3})}{(a^{\nu_1}-a^{\nu_2})(a^{\nu_1}-a^{\nu_3})}, \nonumber\\
	C^2_2 &=& \frac{-(a^{\nu_1}+a^{\nu_3})}{(a^{\nu_2}-a^{\nu_1})(a^{\nu_2}-a^{\nu_3})}, \nonumber\\
	C^2_3 &=& \frac{-(a^{\nu_1}+a^{\nu_2})}{(a^{\nu_3}-a^{\nu_1})(a^{\nu_3}-a^{\nu_2})},\nonumber\\
	C^3_1 &=& \frac{1}{(a^{\nu_1}-a^{\nu_2})(a^{\nu_1}-a^{\nu_3})}, \nonumber\\
	C^3_2 &=& \frac{1}{(a^{\nu_2}-a^{\nu_1})(a^{\nu_2}-a^{\nu_3})}, \nonumber\\
	C^3_3 &=& \frac{1}{(a^{\nu_3}-a^{\nu_1})(a^{\nu_3}-a^{\nu_2})}.\nonumber
\end{eqnarray}
In these solutions we initially assume that all three exponents should be different from each other $\nu_1 \neq \nu_2 \ne \nu_3 \ne \nu_1$. However, as we will see, as in the 2-additive case, the way they are combined to compose the solution, relation (\ref{coefficientsC}), leads to $E^{(N)}(\vec{e})=(C^1_1e_1+C^2_1e_2+C^3_1e_3)N^{\nu_1} + (C^1_2e_1+C^2_2e_2+C^3_2e_3)N^{\nu_2} + (C^1_3e_1+C^2_3e_2+C^3_3e_3)N^{\nu_3}$, for which no divergence appears for $\nu_i=\nu_j$.

The previous results can be sumarized as the following statement.

\textbf{Proposition:} Let $\varrho \in \mathcal{B}\left(\mathcal{H}\right)$ and $N\in \mathds{P}_{a}$. If $\mathcal{E}$ is a 3-additive function such that $\mathcal{E}(\varrho^{\otimes N})=E^{(N)}(\vec{e})=\vec{N}\cdot\vec{e}$, with $\vec{e}=(e_1,e_2,e_3)=(\mathcal{E}(\varrho),\mathcal{E}(\varrho^{\otimes a}), \mathcal{E}(\varrho^{\otimes a^2}))$, where $E^{(a^3)}(\vec{e})=x\text{ }e_1+y\text{ }e_2+z\text{ }e_3$, with $x$, $y$ and $z$ (equivalently $\nu_1$, $\nu_2$ and $\nu_3$) known, then, we have:
\begin{eqnarray*}
E^{(N)}(\vec{e})=\frac{a^{\nu_2+\nu_3}e_1 -(a^{\nu_2}+a^{\nu_3})e_2 +e_3}{(a^{\nu_1}-a^{\nu_2})(a^{\nu_1}-a^{\nu_3})}N^{\nu_1}\nonumber \\ +\frac{a^{\nu_1+\nu_3}e_1 -(a^{\nu_1}+a^{\nu_3})e_2+e_3}{(a^{\nu_2}-a^{\nu_1})(a^{\nu_2} -a^{\nu_3})}N^{\nu_2} \nonumber \\ + \frac{a^{\nu_1+\nu_2}e_1 -(a^{\nu_1} +a^{\nu_2})e_2+e_3}{(a^{\nu_3}-a^{\nu_1})(a^{\nu_3}-a^{\nu_2})}N^{\nu_3},
\end{eqnarray*}
where $\nu_k=\log_a\lambda_k$, $k=1,2,3$, with $\lambda_k$ being the $k$-th eigenvalue of matrix (\ref{qmatrix}).

The 3-additive function normalized per copy, $E^{(N)}(\vec{e})/N$, is unbounded if any of the $\nu_j > 1$ and vanishes if $\nu_1,\nu_2,\nu_3<1$, in the limit of large $N$. Again, we have permutation symmetry throughout, i. e., invariance under $\nu_1 \leftrightarrow \nu_2$, $\nu_2 \leftrightarrow \nu_3$ or $\nu_1 \leftrightarrow \nu_3$. The cases in which two or three $\nu_j$ coincide must be dealt with care. For instance, if we make $\nu_1=\nu$, $\nu_2=\nu+\delta$, and $\nu_3=\nu+\epsilon$, take first the limit $\delta \rightarrow 0$ and afterwards the limit $\epsilon \rightarrow 0$, we get:
\begin{eqnarray}
\nonumber
E^{(N)}(\vec{e})=N^{\nu}\left[ 1-\frac{3}{2}\log_a N+\frac{1}{2}(\log_a N)^2\right]e_1 \\
\nonumber
- \left(\frac{N}{a}\right)^{\nu}\left[2\log_a N+\frac{1}{2}(\log_a N)^2\right]e_2\\
+ \left(\frac{N}{a^2}\right)^{\nu}\left[\frac{1}{2}\log_a N+\frac{1}{2}(\log_a N)^2\right]e_3,
\end{eqnarray}
for $\nu_1=\nu_2=\nu_3$. In particular, for $\nu=1$ and in the limit of large $N$, $E^{(N)}(\vec{e})/N$ diverges as $\sim (\log_a N)^2$.

In the next subsection we use a 3-additive function to describe the one-shot-distillable (OSD) entanglement of a family of isotropic two-qudit states. We assess the effectiveness of our framework by comparison with computational data obtained via linear program reported in \cite{ieee}.
\subsection{Comparison with numeric data on One Shot Distillable (OSD) entanglement} 
\label{isotropic}

In this section we compare an appropriate 3-additive function with the OSD entanglement of a family of symmetric states (see below). The OSD entanglement can be defined for several inequivalent classes of operations as, for instance, Local Operations and Classical Communications (LOCC) and Positive Partial Transpose (PPT) preserving operations. Throughout this work, whenever we mention the OSD entanglement, we are referring to PPT operations. Thus, we do not make this explicit in the notation. The one shot $\varepsilon$-error distillable entanglement of a bipartite state $\varrho$, with respect to PPT operations, is defined as the following optimization \cite{ieee}:
\begin{equation}
E_{OSD}^{\varepsilon}=\log_2 \max \{k \in \mathds{N} | F(\varrho,k)\ge 1-\varepsilon\},
\end{equation}
where $F$ is the fidelity of distillation defined in \cite{rains}, $$F(\varrho,k)=\max_{\tiny \Pi \in PPT} {\rm Tr} [\Pi (\varrho)\Psi_k],$$ $\Psi_k$ being the maximally entangled state of two $k$-dimensional systems: $\Psi_k=|\psi_k\rangle\langle \psi_k|$ ($|\psi_k\rangle=1/\sqrt{k}\sum_{i=0}^k|i\rangle |i\rangle$). The maximization is over all PPT operations.

We will calculate the OSD entanglement of $N$ copies of two entangled $d$-dimensional states $\varrho^{\otimes N}_F$, where
\begin{equation}
\varrho_{F} = F \Psi_d + (1-F)\frac{\mathds{1}-\Psi_d}{d^2-1}, \text{ }\text{ }\text{ with }\text{ }0\leq F\leq 1, 
\label{isostate}
\end{equation} 
$F$ being the fidelity of the state and $\Psi_d$ is the maximally entangled state of two qudits. Although, presently, there is no closed analytical expression for the OSD entanglement of $\varrho_{F}$, the authors of \cite{ieee} showed that the problem can be written as a linear program. 

Superactivation is a common trait of OSD entanglement and will show up for all parameters ($F$ and $d$) we address here, that is, we find $\mathcal{E}_{OSD}(\varrho_F^{\otimes N})=0$ for $N<N_{S.A.}$ and $\mathcal{E}_{OSD}(\varrho_F^{\otimes N})>0$ for $N\ge N_{S.A.}$  We will take this into account, in the framework of 3-additivity, by setting $e_1=0$ and $a=N_{S.A.}$. In addition, in \cite{scal1}, it was demonstrated that the OSD entanglement of a family of Bell states has terms proportional to $N$ and $\sqrt{N}$. Also in this reference, it was shown that the same terms plus an additive constant that corresponds to the asymptotic value of the regularized OSD entanglement, provide a good description of the OSD entanglement of $\varrho_{F}$. For these reasons we use the parameters $(\nu_1,\nu_2,\nu_3)=(1,\frac{1}{2},0)$, where the ordering of the exponents is irrelevant due to the aforementioned permutation symmetry. The regularized expression, then, becomes completely determined and reads
\begin{widetext}
\begin{eqnarray}
\frac{E^{(N)}(\vec{e})}{N}=-\left[\frac{(\sqrt{a}+1)}{(a-\sqrt{a})(a-1)}
+\frac{(a+1)}{(\sqrt{a}-a)(\sqrt{a}-1)\sqrt{N}}+
\frac{(a +\sqrt{a})}{(1-a)(1-\sqrt{a})N}\right]e_2 \nonumber\\
+\left[\frac{1}{(a-\sqrt{a})(a-1)}+\frac{1}{(\sqrt{a}-a)(\sqrt{a}-1)\sqrt{N}}+ \frac{1}{(1-a)(1-\sqrt{a})N}\right]e_3, \label{3fit}
\end{eqnarray}
\end{widetext}
with $e_2=\mathcal{E}(\varrho^{\otimes a})$ and $e_3=\mathcal{E}(\varrho^{\otimes a^2})$. The, finite, regularized asymptotic limit becomes simply:
\begin{equation}
\lim_{N \rightarrow \infty} \frac{E^{(N)}(\vec{e})}{N} = \frac{-(\sqrt{a}+1)e_2+e_3}{(a-\sqrt{a})(a-1)}. 
\label{assuper3}
\end{equation}

We will test expression (\ref{3fit}) for entangled qubits, qutrits and ququarts, where, in all cases, we set the error tolerance to $\varepsilon=0.001$, as in \cite{ieee}. We stress that,
although our formal statements only refer to powers of a given integer $a$, the results below indicate that the 3-additive expression serves as a good approximation to the OSD entanglement of arbitrary $N$. 
\begin{figure}[h]
		\includegraphics[height=5cm,angle=0]{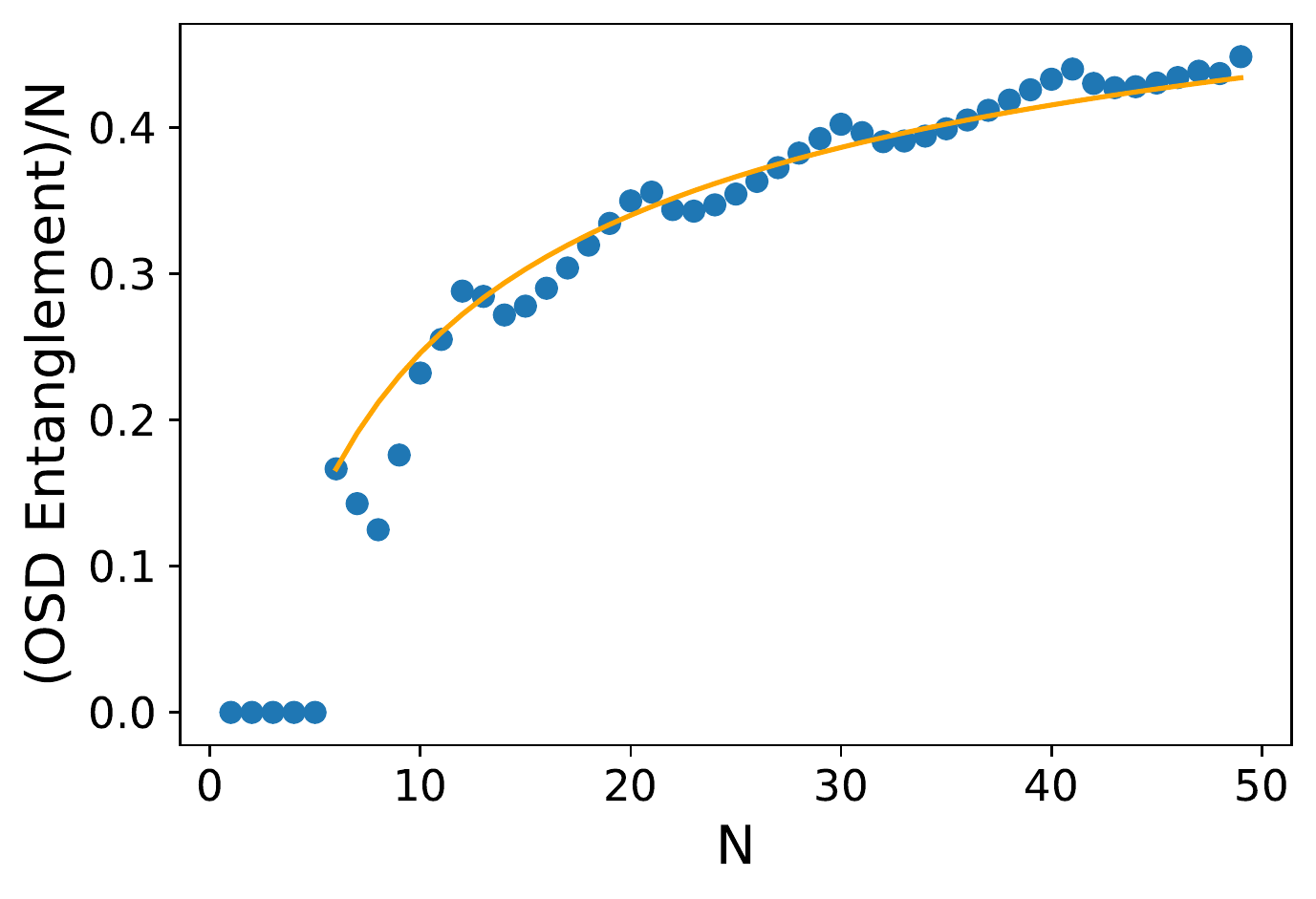}
		\caption{The regularized 3-additive function of Eq. (\ref{3fit}) for two entangled qubits ($d=2$) with $F=0.96$ (superactivation at $a=6$) is shown as a continuous curve to facilitate visualization. The bullets represent the numeric results obtained via linear programming, up to $N=50$.}\label{fig1} 
\end{figure}

We start with the simplest scenario of two entangled qubits ($d=2$), with $F=0.96$. Despite the high fidelity, superactivation occurs only at $N=6=a$ copies. The input parameters, obtained from the mentioned linear program, are $e_2={\cal E}(\varrho^{\otimes 6})=1$ and $e_3={\cal E}(\varrho^{\otimes 36})=36 \times 0.405$. In figure \ref{fig1}, 
the corresponding numeric results (bullets) are shown, up to $N=50$, and compared with the corresponding 3-additive function, equation (\ref{3fit}).

\begin{figure}[h]
		\includegraphics[height=5cm,angle=0]{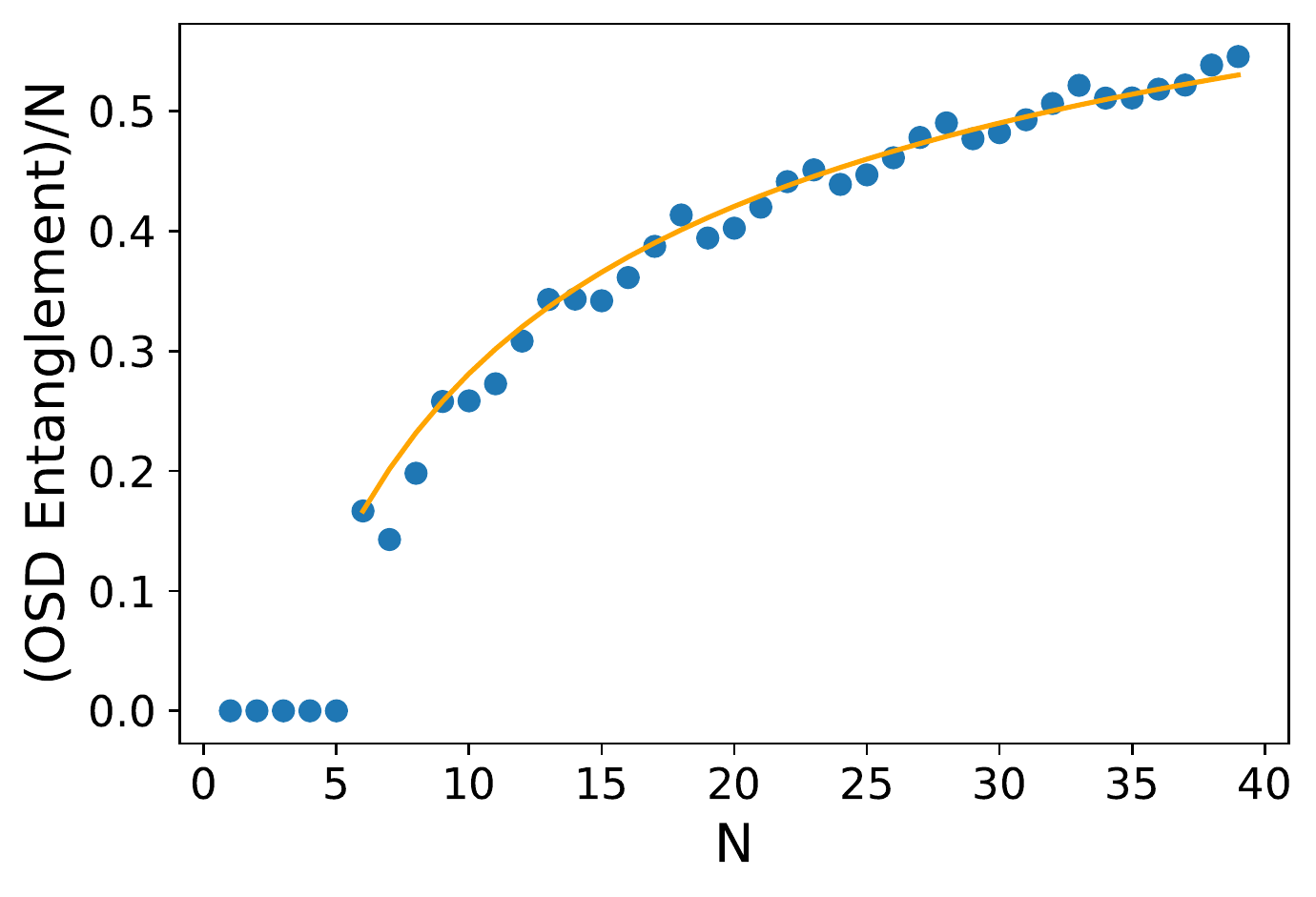}
		\caption{The regularized 3-additive function of Eq. (\ref{3fit}) for two entangled qutrits ($d=3$) with $F=0.9$ (superactivation at $a=6$) is shown as a continuous curve to facilitate visualization. The bullets represent the numeric results obtained via linear programming, up to $N=40$.} \label{fig2}
\end{figure}

Next we address, in more detail, the exact system and parameters used in \cite{ieee}: two entangled qutrits ($d=3$) with $F=0.9$.
With these parameters, the OSD entanglement is also superactivated for $a=6$ copies. The input parameters are $e_2={\cal E}(\varrho^{\otimes 6})=1$ and $e_3={\cal E}(\varrho^{\otimes 36})=36 \times 0.518$. We reproduce the numeric results of \cite{ieee} up to $N=40$ and compare them to the corresponding 3-additive function in figure \ref{fig2}. 
In addition, the asymptotic limit given by the simple expression (\ref{assuper3}) is $\approx 0.86$, while the actual result, determined in \cite{ieee} is $\approx 0.81$, an accuracy of $\sim 94\%$. 
Note that, as it was discussed in the introduction, the asymptotic regime may become dominant only for very high $N$. In the present case, for $N=40$ asymptotic expressions would not be a good description, since the points still do not show a tendency to saturation. Indeed, for $N=40$, the OSD entanglement per copy is bellow $0.55$, while the approximate saturation should occur around $0.81$. This feature is qualitatively valid for the other cases studied here.
%

Finally, we go to higher dimensional states: $d=4$ with $F=0.9$ (superactivation for 5 copies). The input parameters are $e_2={\cal E}(\varrho^{\otimes 5})=1$ and $e_3={\cal E}(\varrho^{\otimes 25})=25 \times 0.659$. We use the linear program given in \cite{ieee} to calculate the OSD entanglement, up to $N=30$. Again, the continuous line represents expression (\ref{3fit}). Although for $N \sim 30$ we have $E^{(N)}(\vec{e})/N \approx 0.7$, the points do not show a tendency to saturate. Indeed, the asymptotic value of $E^{(N)}(\vec{e})/N$ is around $1.197$, as predicted by (\ref{assuper3}). Differently from the case $d=3$ (and obviously from the case $d=2$), for $d=4$ and $F=0.9$ more than a singlet can be distilled per ququart, asymptotically, according to the 3-additive formula.
\begin{figure}[h]
		\includegraphics[height=5cm,angle=0]{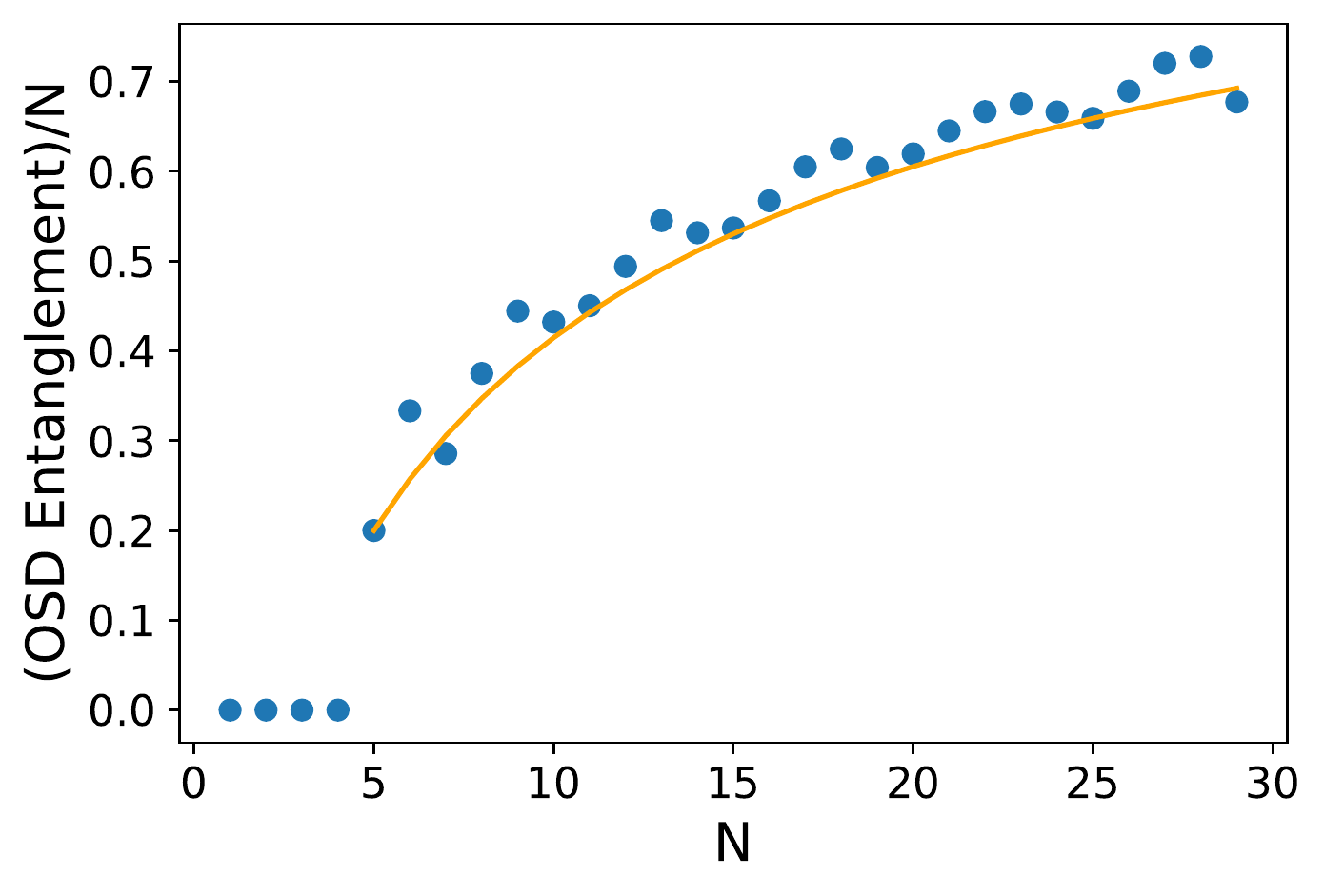}
		\caption{The regularized 3-additive function of Eq. (\ref{3fit}) for two entangled ququarts ($d=4$) with $F=0.9$ (superactivation at $a=5$) is shown as a continuous curve to facilitate visualization. The bullets represent the numeric results obtained via linear programming, up to $N=30$.} 
\end{figure}

It is clear from these examples that the simple 3-additive expression provides a good qualitative description, with a quantitative accuracy which is typically above 90$\%$. We stress that there is no free fitting parameter to be adjusted in formula (\ref{3fit}). 


It is worth mentioning that several internal consistency condition can be derived. For instance, Eq. (\ref{assuper3}) and the positivity of ${\cal E}$ imply:
\begin{equation}
e_3\ge (\sqrt{a}+1) e_2.
\end{equation}
If the numeric determination of $e_2$ and $e_3$ leads to a violation of the above condition, equation (\ref{3fit}) can be dismissed as a possible description of the figure of merit under
investigation. It is clear that this corresponds to a nontrivial constraint only for sub-additive quantifiers [${\cal E}(\varrho^{\otimes N})/N$ being a decreasing function of $N$].

\section{Closing remarks}
For several resource quantifiers whose domains are high-dimensional Hilbert-Schmidt spaces,  ${\cal E}(\varrho^{\otimes N}): {\cal B(H)}^{\otimes N} \mapsto \mathds{R}_{+}$, there is a ``gray zone'' between small values of $N$ and the asymptotic regime $N \rightarrow \infty$, for which the evaluation of  ${\cal E}$ becomes a prohibitive task. 
In this regard, any method that helps to circumvent the direct evaluation of these figures of merit, avoiding or reducing the need of optimization processes, may be useful in quantum information science. In this work we have introduced the concept of $q$-additivity by considering a natural extension of the notion of additive measures, thus, going from ${\cal E}(\varrho^{\otimes N})=N e$ to ${\cal E}(\varrho^{\otimes N})=\vec{N} \cdot \vec{e}$. 

The formalism naturally accommodates the phenomenon of superactivation, by setting one of the components of $\vec{e}$ to zero. Although we derived explicit expressions for 2- and 3-additive functions only, the general framework of $q$-additivity can be addressed via linear systems of equations. Therefore, the problem is amenable to computational treatment for larger values of $q$. It is hoped that this framework can provide a useful tool to deal with the mentioned gray-zone problem.

Regarding the usefulness prospects of the presented formalism, in a broader scenario, a key question is: Given a resource quantifier ${\cal E}$, can one prove (or disprove) $q$-additivity (or more generally $q$-scalability \cite{scal1}) directly from the definition of ${\cal E}$, without actually calculating ${\cal E}$? A positive answer for a given quantifier, at least for a particular class of states, would mean a huge simplification.
Another research direction that may be interesting is the possibility to define computable measures by using $q$-additivity as an ingredient, from the outset. 

We mostly referred to entanglement measures. The presented results, however, are equally valid for coherence measures \cite{coh,coh2,coh3,coh4,coh5,coh6,coherence}, for instance, and for the evaluation of other quantum figures of merit.

\begin{acknowledgments}
This work received financial support from the Brazilian agencies Coordena\c{c}\~ao de Aperfei\c{c}oamento de Pessoal de N\'{\i}vel Superior (CAPES), Funda\c{c}\~ao de Amparo \`a Ci\^encia e Tecnologia do Estado de Pernambuco (FACEPE), and Conselho Nacional de Desenvolvimento Cient\'{\i}fico  e Tecnol\'ogico through its program CNPq INCT-IQ (Grant 465469/2014-0).
\end{acknowledgments}


\begin{thebibliography}{99}
\bibitem{comm0} The term ``resource'' is employed in the general sense, since the results to be derived do not rely on all 
the requirements for a quantity to be a resource (in the resource-theoretic sense).
\end{thebibliography}
\end{document}